\documentstyle[preprint,aps]{revtex}
\begin{document}
\draft
\title{ Electromagnetic waves in NUT space: Solutions to
the Maxwell equations}
\author{
 Mohammad Nouri-Zonoz
\thanks{Email: nouri@khayam.ut.ac.ir}}
\address{Physics Department, University of Tehran, 14352, Tehran
, Iran.
\\
Institute for Studies in Theoretical Physics and Mathematics,
Farmanieh, 19395-5531, Tehran, Iran.
\\{and}\\
Institute of Astronomy, Madingley Road, Cambridge, CB3 0HA, UK.}
\maketitle
\begin{abstract}
In this paper, using the Newman-Penrose formalism, we find the
Maxwell equations in NUT space and after separation into angular
and radial components solve them analytically. All the angular equations
 are solved in terms of Jaccobi polynomials. The radial equations are 
 transformed into Hypergeometric and Heun's equations with  the 
 right hand sides including terms of different order in the frequency of the 
perturbation which allow solutions in the expansion of this parameter.

\end{abstract}

\section{Introduction}
An effective way to understand and analyse characteristics of an 
spacetime is to study its behaviour under different kind of perturbations
including the most studied one {\it electromagnetic perturbations}. Indeed 
interest in the analytic solutions of Maxwell's
equations in curved spacetimes, specially in the well known solutions of Einstein
field equations, arose from the study of stability of those spacetimes under electromagnetic perturbations. On the other hand the
(Taub)-NUT solution of Einstein vacuum field equations [1,2] has found
importance both in the discussions on magnetic monopoles in gauge  
theories [3-4] and in
string theory [5-6], though by itself interpreted as the metric of a
mass M endowed with a gravitomagnetic monopole with strength $l$,
the so called NUT parameter or {\it magnetic mass} [7]. Therefore it 
seems appropriate to study
electromagnetic waves in NUT space through the solutions of the
Maxwell equations in this spacetime and use the results
to obtain a deeper understanding of the physical aspects of NUT
space . \\
In a recent paper [8] perturbations of the Kerr-Taub-NUT spacetime 
by massless fields of spin $s \le 2$ are studied by introducing a 
 master equation describing these fields. The authors have also 
discussed the coupling between the gravitomagnetic monopole moment $l$
and the spin of the perturbation field in the context of 
gravitomagnetism. In [9] Klein-Gordon and Dirac equations in Kerr-NUT space have been studied with an emphasis on a duality transformation involving the exchange of mass and NUT parameters. It is shown that the same duality transformation also exchanges the scalar and spinor perturbations.  
In the present article, as stated in the abstract, we have a more  restricted task in front of us 
and that is to study electromagnetic perturbations (massless spin 1 field) 
in NUT space (and not Taub-NUT space) but we intend here to solve the 
resulted equations more or less anlytically. 

To do so we start with the fact that as in the case of the Kerr metric, 
because NUT space is
stationary and axisymmetric \footnote{For a review on the physical symmetries of NUT space refer to [7] and [10].}, one could express a general solution to the
Maxwell equations in this space as a superposition of different
modes with a time- and $\phi$-dependence given by [11]
$$e^{i(\omega t + m\phi)}\eqno(1)$$
where $m=0,\pm 1,\pm 2, ....$ and $\omega$ is the frequency of
each mode. Treating the electromagnetic field as a perturbation,
it is natural to think of $\omega$ as a small parameter in terms
of which the appropriate functions could be expanded if necessary.
 
\section{Maxwell equations in NUT space}
In this section using 
Newman-Penrose formalism [12] we
separate the remaining two variables, $r$ and $\theta$ and obtain the 
radial and angular equations involved.
To do so we start with the NUT metric given by the line element
$$ds^2=f(r)\left( dt-2l {\rm cos}\theta d\phi\right)^2 - {1\over
f(r)}dr^2 -(r^2+l^2)(d\theta^2 + {\rm sin}^2\theta d\phi^2)\eqno(2)$$
where $f(r)=1-{2(Mr+l^2)\over r^2+l^2} > 0$ and $l$ is called the
{\it magnetic mass} or NUT factor. Using the tangent vectors to the null 
geodesics in NUT space, one can easily show that the following
set of null vectors form a null-tetrad basis for the above metric
$$l^\mu =({1\over f(r)},1,0,0)\;\;\;\;\;\;   ,   \;\;\;\;\;\;\;\;\; l_\mu =
(1,-{1\over f(r)},0, -2l{\cos}\theta)$$
$$n^\mu = ({1\over 2}, -{f(r)\over 2},0,0)\;\;\;\;\;\;   , \;\;\;\;\;\;\;\;
n_\mu = ({f(r)\over 2}, {1\over 2}, 0, -l{\cos}\theta f(r))$$
$$m^\mu = {1\over \sqrt{2(r^2+l^2)}}(2il{\rm cot}\theta , 0, 1,
{\rm i}\csc\theta)\;\;\;\;\;\;   ,   \;\;\;\;\;\;\;\;\; m_\mu =
\sqrt{{r^2+l^2\over 2}}(0, 0, -1, -{\rm i}\sin \theta)$$
$$\bar{m}^\mu = {1\over \sqrt{2(r^2+l^2)}}(-2{\rm i}l{\rm cot}\theta , 0, 1,
-{\rm i} \csc\theta)\;\;\;\;\;\; , \;\;\;\;\;\;\;\;\; \bar{m}_\mu =
\sqrt{{r^2+l^2\over 2}}(0, 0, -1, {\rm i} \sin\theta)\eqno(3)$$
Using the above null tetrad one can find the following spin coefficients
of NUT metric;
$$\kappa = 0 , \;\; \sigma =0 ,\;\; \lambda =0 , \;\; \nu =0 ,
\;\; \tau =0 , \;\; \pi =0$$
$$\rho = {-r + {\rm i}l \over r^2 + l^2}\;\;\;\; ; \;\;\;\; \mu = {f(r)\over 2}({{-r + {\rm i}l} \over r^2 + l^2})$$
$$\epsilon = {{\rm i}l \over 2(r^2+ l^2)}\;\;\;\; ; \;\;\;\; \gamma = {1\over 4} \left( f^\prime(r) + {{\rm i}lf(r)\over r^2 + l^2}\right)$$
$$\alpha = {-1\over 2} {{\rm cot}\theta \over \sqrt{2(r^2 + l^2)}}\;\;\;\; ; \;\;\;\; \beta = {1\over 2} {{\rm cot}\theta \over \sqrt{2(r^2 + l^2)}}\eqno(4)$$
Having found the spin coefficients of NUT metric we can now write
the Maxwell equations in this space using the Newman-Penrose
formalism [12]. For this we replace the antisymmetric Maxwell
tensor $F_{\mu\nu}$ with the following three {\it complex} scalars
[11];
$$\phi_0 = F_{13} = F_{\mu\nu}l^\mu m^\nu$$
$$\phi_1 = \frac{1}{2} (F_{12} + F_{43}) = \frac{1}{2} F_{\mu\nu}(l^\mu n^\nu +
\bar{m}^\mu m^\nu)$$
$$\phi_2 = F_{42} = F_{\mu\nu} \bar{m}^\mu n^\nu\eqno(5)$$where
$F_{ij}$ ($i,j=1,2,3,4$)  and $F_{\mu\nu}$ ($\mu,\nu=0,1,2,3$) are the components of the Maxwell
tensor in the tetrad and tensor bases respectively. One can invert equations (5) to find
$$F_{\mu\nu}=2\left[\phi_0 \bar{m}_{[\mu}n_{\nu ]} + \phi_1 (n_{[\mu}l_{\nu ]} + m_{[\mu}\bar{m}_{\nu
]})+\phi_2 l_{[\mu}m_{\nu ]}\right] + {C \;.\; C}$$
In terms of the above scalars the Maxwell equations
become;
$$ D \phi_1 - \delta^* \phi_0 = (\pi - 2\alpha) \phi_0 + 2\rho
\phi_1 - \kappa \phi_2$$
$$D \phi_2 - \delta^* \phi_1 = - \lambda \phi_0 +
2 \pi \phi_1 + (\rho -2\epsilon )\phi_2$$
$$ \delta \phi_1 - {\cal P} \phi_0 = (\mu -2\gamma)\phi_0 + 2\tau
\phi_1 -\sigma \phi_2$$
$$\delta \phi_2 - {\cal P} \phi_1 = -\nu \phi_0 + 2\mu
\phi_1 + (\tau -2\beta) \phi_2 \eqno(6)$$
Where in the above equations $ D, {\cal P}, \delta$ and $\delta^*$
are the special symbols for the basis vectors ${\bf l}, {\bf n}, {\bf m}$  and $\bar{\bf m}$
when they are considered as the directional derivatives. For NUT
space these are;
$${\bf l} \equiv D ={1\over f(r)} {\partial \over \partial t} + {\partial \over \partial r}$$
$${\bf n}\equiv {\cal P} = {1 \over 2}{\partial \over \partial t}  - {f \over 2}{\partial \over \partial r} $$
$${\bf m}\equiv \delta = {1\over \sqrt{2(r^2+l^2)}}(2{\rm i}l{\rm
cot}\theta {\partial \over \partial t} +  {\partial \over \partial \theta} +
{{\rm i}\over {\rm sin}\theta} {\partial \over \partial \phi})$$
$$\bar{\bf m} \equiv \delta^* = {1\over \sqrt{2(r^2+l^2)}}(-2{\rm i}l{\rm
cot}\theta {\partial \over \partial t} + {\partial \over \partial \theta} -
{{\rm i}\over {\rm sin}\theta}{\partial \over \partial \phi} )\eqno(7)$$
Now if we use the general form (1) for the $(t,\phi)$- dependence of
the wave, the above operators take the following forms;
$$D = \partial_r + {\rm i} {\omega \over f(r)}$$
$${\cal P} = {{\rm i} {\omega}\over 2}- {f(r)\over 2}\partial_r = - {f(r)\over
2}D^\dag$$
$$\delta = {1\over \sqrt{2(r^2+l^2)}}(-2l \omega {\cot}\theta +
\partial_\theta -m {\csc}\theta)$$
$$\delta^* = {1\over \sqrt{2(r^2+l^2)}}(2l \omega {\cot}\theta +
\partial_\theta +m {\csc}\theta)\eqno(8)$$
Now we define the above operators in the following way;
$${\bf l}=D={\cal D}_0 \;\;\;\;\;\;\; , \;\;\;\;\;\;\; {\bf n}={\cal P} = - {f(r)\over
2}{\cal D}_0^\dag$$
$${\bf m}=\delta = {1\over \sqrt{2(r^2+l^2)}}{\cal L}_0^\dag \;\;\;\;\;\;\; ,
 \;\;\;\;\;\;\; {\bar {\bf m}} = \delta^* = {1\over \sqrt{2(r^2+l^2)}}{\cal L}_0$$
where
$${\cal D}_n = \partial_r + {\rm i} {\omega \over f(r)} + 2n {r-M
\over \Delta}$$
$${\cal D}_n^\dag = \partial_r - {\rm i} {\omega \over f(r)} + 2n
{r-M\over \Delta}$$
$${\cal L}_n = \partial_\theta + Q + n{\cot}\theta$$
$${\cal L}_n^\dag = \partial_\theta - Q + n{\cot}\theta\eqno(9)$$
in which $\Delta =r^2 -2Mr - l^2 $ and $Q = 2l\omega {\cot}\theta + m{
\csc}\theta$. We note that ${\cal L}_n$ is (up to a factor ${1\over \sqrt{2(r^2+l^2)}}$) nothing but one of the weighted derivative operators  
introduced in the compacted-spin coefficient formalism [13].
Now substituting the above operators in the Maxwell equations (6) we
have;
$${1\over \sqrt{2(r^2 + l^2)}} {\cal L}_1 \phi_0 =
 ({\cal D}_0 + 2{r-{\rm i}l \over r^2 + l^2})\phi_1$$
$${1\over \sqrt{2(r^2 + l^2)}} {\cal L}_0 \phi_1 =
 ({\cal D}_0 + {r \over r^2 + l^2})\phi_2$$
$${1\over \sqrt{2(r^2 + l^2)}} {\cal L}_1^\dag \phi_2 =
-{f(r)\over 2}({\cal D}_0^\dag + 2{r-{\rm i}l \over r^2 + l^2})\phi_1$$
$${1\over \sqrt{2(r^2 + l^2)}} {\cal L}_0^\dag \phi_1 =
 -{f(r)\over 2}({\cal D}_1^\dag - {r \over r^2 + l^2})\phi_0$$
The above equations will look more compact if we change the  variables to;
$$\Phi_0 = \phi_0 \;\;\;\;\; , \;\;\;\;\; \Phi_1 =
\sqrt{2(r^2 + l^2)}\phi_1 \;\;\;\;\; , \;\;\;\;\;\Phi_2 =
{2(r^2 + l^2)}\phi_2$$ where now, using the definition of the
operator ${\cal D}_n$, we have;
$${\cal L}_1 \Phi_0 = ({\cal D}_0 + {r-2{\rm i}l \over r^2 + l^2})
\Phi_1\eqno(10)$$
$${\cal L}_0 \Phi_1 = ({\cal D}_0 - {r \over r^2 + l^2})\Phi_2\eqno(11)$$
$${\cal L}_1^\dag \Phi_2 = -\Delta({\cal D}_0^\dag + {r-2{\rm i}l \over r^2 + l^2})
\Phi_1\eqno(12)$$
$${\cal L}_0^\dag \Phi_1 = -\Delta({\cal D}_1^\dag - {r \over r^2 + l^2})
\Phi_0\eqno(13)$$

The commutativity of the operators ${\cal L}_0^\dag$ and $-\Delta({\cal D}_0 +
{r-2{\rm i}l\over r^2 +l^2})$ enables us to eliminate $\Phi_1$ from equations (10)
and (13) and get;
$$\left[{\cal L}_0^\dag {\cal L}_1 + \Delta ({\cal D}_1 + {r-2{\rm i}l\over r^2
+l^2})({\cal D}_1^\dag  - {r\over r^2+l^2})\right]\Phi_0(r,\theta)=0 \eqno(14)$$
where we have used the fact that ${\cal D}_n \Delta = \Delta {\cal
D}_{n+1}$.
Similarly the commutativity of the operators ${\cal L}_0$ and
$-\Delta({\cal D}_0^\dag + {r-2{\rm i}l\over r^2 +l^2})$
will allow us to eliminate $\Phi_1$ from equations (11) and (12) and get;
$$\left[{\cal L}_0 {\cal L}_1^\dag + \Delta ({\cal D}_0^\dag + {r-2{\rm i}l\over r^2
+l^2})({\cal D}_0 - {r\over r^2+l^2})\right]\Phi_2(r,\theta)=0
\eqno(15)$$ Now our task is to find solutions to the equations (14)
and (15) by separating them into angular and radial parts. This
separation can be achieved by choosing the following general form
for $\Phi_0(r,\theta)$ and $\Phi_2(r,\theta)$;
$$\Phi_0(r,\theta) = {\cal R}_0(r) \Theta_0 (\theta)\;\;\;\;\;\;\;
 {\rm and} \;\;\;\;\;\;\;
\Phi_2(r,\theta) = {\cal R}_2(r) \Theta_2 (\theta)\eqno(16)$$
Substituting (16) in (14) and (15) we have;
$${\cal L}_0^\dag {\cal L}_1 \Theta_0(\theta) = (\partial_\theta -
Q)(\partial_\theta + Q + {\rm cot}\theta )\Theta_0 (\theta)=
-\xi \Theta_0 (\theta)\eqno(17a)$$
$$\Delta ({\cal D}_1 + {r-2{\rm i}l\over r^2+l^2})
({\cal D}_1^\dag- {r\over r^2+l^2}){\cal R}_0(r) = \xi {\cal R}_0\eqno(17b)$$
$${\cal L}_0 {\cal L}_1^\dag \Theta_2(\theta) = (\partial_\theta +
Q)(\partial_\theta - Q + {\rm cot}\theta )\Theta_2 (\theta)=
-\xi \Theta_2 (\theta)\eqno(18a)$$
$$\Delta ({\cal D}_0^\dag + {r-2{\rm i}l\over r^2+l^2})
({\cal D}_0- {r\over r^2+l^2}){\cal R}_2(r) = \xi {\cal
R}_2\eqno(18b)$$ where $\xi $ is the separation constant. These are
analogous to the Teukolsky equations in Kerr space [11,14]. Note that
we have not distinguished between the separation constants that
derived from equations (14) and (15) and the reason is that
equations (17a) and (18a) determine the same set of eigenvalues
$\xi$. This can be seen by considering equation (17a) where, apart
from the regularity requirement of $\Theta_0(\theta)$ at $\theta=0$
and $\theta=\pi$ for the determination of $\xi$, the operator
acting on $\Theta_2(\theta)$ in equation (18a) is the same as the
operator acting on $\Theta_0$ in (17a) if we replace $\theta$ by
$\pi - \theta$ and this is the case because the general relation
${\cal L}_n(\theta)= -{\cal L}_n^\dag(\pi - \theta)$ holds. So a
proper solution , $\Theta_0(\theta;\xi)$, of equation (17a) with
eigenvalue $\xi$ , is also a solution of equation (18a) with the
same eigenvalue, if we replace $\theta$ by $\pi - \theta $ in
$\Theta_0(\theta;\xi)$.
\section{Solution to the angular equation}
In this section we solve equation (17a) analytically. It is
obvious from the form of the equation (18a) and the discussion
above that it will have the same set of eigenvalues as (17a).
First of all we rewrite equation (17a) in the following form;
$$(\partial^2_\theta  + {\rm cot}\theta \partial_\theta
)\Theta_0(\theta) + \Theta_0(\theta) \partial_\theta (Q+ {\rm
cot}\theta) -(Q^2+Q {\rm cot}\theta -\xi)\Theta_0(\theta) = 0$$
Substituting for $Q$ from its definition, the above equation
becomes;
\begin{eqnarray*}
\lefteqn{(\partial^2_\theta  + {\rm cot}\theta \partial_\theta
)\Theta_0(\theta) - {1\over {\rm sin}^2\theta}(2l\omega +
1)\Theta_0(\theta)-} \hspace{1.25in} \\
& &   \left( {2m{\cos}\theta \over {\rm sin}^2\theta}+
4l^2\omega^2 {\rm cot}^2\theta + {m^2 \over {\rm sin}^2\theta} +
4l\omega m {{\cos}\theta \over {\rm sin}^2\theta} + 2l \omega
{\rm cot}^2\theta - \xi \right)\Theta_0(\theta)=0
\end{eqnarray*}
Now changing the variable to $x={\cos}\theta$ we have;
$$\left[ (1-x^2){{\rm d}^2 \over {\rm dx}^2} - 2x {{\rm d} \over {\rm
d x}}- {(2mx + 4b^2x^2 + m^2 + 4bmx + 2bx^2 + 2b + 1) \over 1-x^2}
+\xi \right]\Theta_0(x) =0 \eqno(19)$$ where $b = l\omega$.
Dominant behaviour of the eigenfunctions of the above equation at
singular points $x=\pm 1$ are given by $(1 \mp x)^{|1+2b\pm m
|\over 2}$ respectively. Introducing $\alpha = |1+2b+m|$ and $\beta
= |1+2b -m |$, and substituting for $\Theta_0$ the following
expression;
$$({1-x \over 2})^{\alpha /2} ({1+x \over 2})^{\beta /2} U(x)$$
in (19) it will be transformed into;
$$(1-x^2){{\rm d}^2 U(x) \over {\rm dx}^2} + [\beta - \alpha
-(2+\alpha + \beta)x ]{{\rm d}U(x) \over {\rm dx}}+[\xi - ({\alpha
+\beta \over 2})({\alpha+\beta \over 2}+1) + 4b^2+2b]U(x)=0\eqno(20a)$$ 
This is the differential equation satisfied by  the Jacobi polynomials [15,16]
$$P_n^{(\alpha,\beta)}(x) = {(-1)^n \over 2^n n!} (1-x)^{-\alpha} (1+x)^{-\beta} {{\rm d}^n \over {\rm d}x^n}\left[ (1-x)^{\alpha+n}(1+x)^{\beta+n} \right]$$
 if we  identify;
$$\xi + 4b^2 + 2b = [n+({\alpha+\beta \over 2})][n+({\alpha+\beta \over
2}+1)]=j(j+1)\eqno(20b)$$
where $j=n+({\alpha+\beta \over 2})$. So $\Theta_0$ could be written explicitly as follows
$$\Theta_0(x) = {(-1)^n \over 2^{n-{\alpha\over 2}-{\beta\over 2}} n!} (1-x)^{-{\alpha\over 2}} (1+x)^{-{\beta\over 2}} {{\rm d}^n \over {\rm d}x^n}
\left[ (1-x)^{\alpha+n}(1+x)^{\beta+n} \right]\eqno(20c)$$
\section{Solutions to the radial equations}
Now we try to solve the radial equations (17b) and (18b) by
replacing for the operators ${\cal D}$ and ${\cal D}_1$ and
their complex conjugates from (9);
\begin{eqnarray*}
\lefteqn{{\partial^2_r {\cal R}_0 + ({4[r-M]\over \Delta}-
2{{\rm i} l\over r^2+l^2})\partial_r {\cal R}_0} +\left( {\omega^2
(r^2 + l^2)^2-2{\rm i} \omega M (l^2 -r^2) +4{\rm i}r \omega l^2
 \over \Delta^2}\right){\cal R}_0 \hspace{1.5in} }\\
\\ & &  + \left({2-2{\rm i} \omega (r-{\rm i}l)
-\xi \over \Delta} -{4{\rm i} l(r-M)\over \Delta (r^2+l^2)} +
{2{\rm i} l r - l^2\over (r^2+l^2)^2} \right){\cal R}_0 = 0
\hspace{2.2in}(21a)
\end{eqnarray*}

\begin{eqnarray*}
\lefteqn{\partial^2_r {\cal R}_2 - 2{{\rm i} l\over r^2+l^2}\partial_r
{\cal R}_2 \hspace{1.5in} }\\
\\ & &  + \left( {2{\rm i} \omega M (l^2 -r^2) - 4{\rm i}r \omega
l^2 + \omega^2 (r^2 + l^2)^2\over \Delta^2} + {2{\rm i} \omega (r-
{\rm i}l)-\xi \over \Delta} + {2{\rm i} l r - l^2\over (r^2+
l^2)^2}\right){\cal R}_2 =0 \hspace{0.6in}(21b)
\end{eqnarray*}
where $\Delta = r^2-2Mr -l^2 = (r-r_+)(r-r_-)$ with $r\pm = M\pm
\sqrt{M^2+l^2}$ and $r_+$ being the position of the horizon. Therefore in
our discussion below we always have $r \geq r_+$.  
\subsection{Solution to the equation (21b)}
We start by solving equation (21b) which looks simpler. This
equation has four regular singularities at $r=r_{\pm}$ and $r=\pm
{\rm i}l$ and an irregular singularity at $r=\infty$. To remove the
singularities at $r=\pm {\rm i}l$ we use the following
transformation ;

$${\cal R}_2(r) = ({r-{\rm i}l\over r+{\rm i}l})^{1/2} {\cal V}_2(r)\eqno(22a)$$
so that equation (21b) becomes (after some manipulation);
$$\partial^2_r {\cal V}_2(r) + \left( {A\over (r-r_+)^2} + {B\over (r-r_-)^2}
+{C\over (r-r_+)(r-r_-)} + D ({1\over r-r_+} + {1\over r-r_-}) +
\omega^2 \right){\cal V}_2(r)=0 \eqno(22b)$$ where
$$A = r_+ (2M\omega^2 - {\rm i}\omega) + b^2 $$
$$B = r_- (2M\omega^2 - {\rm i}\omega) + b^2 $$
$$C = 2M (2M\omega^2 + {\rm i}\omega) + 2(b^2+b) - \xi$$
$$ D = 2M\omega^2 + {\rm i}\omega \eqno(23)$$
Now if we define
$$r-r_+ = -\varepsilon x \;\;\;\;\;\;\;\;\; \& \;\;\;\;\;\;\;\;\; r-r_- =
\varepsilon (1-x) \eqno(24)$$
where $\varepsilon = 2(M^2+l^2)^{1/2}$ and take $${\cal V}_2 = (-x)^\alpha
(1-x)^\beta \bar{\cal V}_2 $$ (where $\alpha$ and $\beta$ are
constants to be determined later) then equation (22b) will transform
into;
\begin{eqnarray*}
\lefteqn{\partial^2_x \bar{\cal V}_2(x) +2({\alpha \over x}- {\beta \over 1-x})
\partial_x \bar{\cal V}_2(x) +\left[{A\over x^2} + {B\over (1-x)^2}
-{C\over x(1-x)}\right]\bar{\cal V}_2(x)  \hspace{1.5in} }\\
\\ & & + \left[ \varepsilon D ({1\over 1-x} - {1\over x}) +
\varepsilon^2 \omega^2 +
{\alpha (\alpha -1) \over x^2} - {2\alpha \beta \over x(1-x)} +
{\beta (\beta -1) \over (1-x)^2} \right]\bar{\cal V}_2(x) =0
\hspace{1 in}(25)
\end{eqnarray*}
Now in order to eliminate the terms proportional to $1\over x^2$
and $1\over (1-x)^2$, $\alpha$ and $\beta$ should take one of the
following values. ;
$$\alpha_{\pm} = {1\over 2}(1 \pm \sqrt{1-4A})\;\;\;\;\;\;\; , \;\;\;\;\;\;\;
\beta_{\pm} = {1\over 2}(1 \pm \sqrt{1-4B})\eqno(26)$$ As in the
case of the Kerr spacetime one can show that different choices of
the above parameters correspond to solutions which satisfy the
incoming and outgoing boundary conditions [17]. For example
choosing the pair $(\alpha_{-}, \beta_{+})$ corresponds to the
solution satisfying the incoming boundary condition\footnote {
This could also be seen by the fact that $a=0$ limit of the Kerr case and $l=0$
limit of NUT case should both reduce to the electromagnetic perturbation of schwarzschild spacetime.}. With the
above choice of $\alpha$ and $\beta$ equation (25) reduces to;
\begin{eqnarray*}
\lefteqn{\partial^2_x \bar{\cal V}_2(x) +2({\alpha \over x}- {\beta \over
1-x})\partial_x \bar{\cal V}_2(x)+ \left[\varepsilon D
({1\over 1-x} - {1\over x}) \right]\bar{\cal V}_2(x)-
\hspace{1.5in} }\\& & \left[{C\over x(1-x)} +\varepsilon^2 \omega^2 -
{2\alpha \beta \over x(1-x)}\right]\bar{\cal V}_2(x) =0 \hspace{3.3in}\; (27)
\end{eqnarray*}
Now choosing $\bar{\cal V}_2(x)= \exp (i \varepsilon \omega x) \tilde{\cal
V}_1(x)$ the above equation can be written in the following form;
\begin{eqnarray*}
\lefteqn {x(1-x)\partial^2_x \tilde{\cal V}_2(x)+[2\alpha -
2(\alpha + \beta)x ]\partial_x \tilde{\cal V}_2(x) - p q
\tilde{\cal V}_2(x)= \hspace{1.5in} }\\ & & ((2M+\varepsilon)(2M\omega^2 +
{\rm i}\omega)+2(3l^2\omega^2+2l\omega)-j(j+1)+ 2\alpha
\beta-2{\rm i}\varepsilon \omega \alpha) \tilde{\cal V}_2(x)
 + \hspace{1.5in} \\ & & ( 2{\rm i}\varepsilon \omega [\alpha+\beta] -
 2\varepsilon (2M\omega^2 + {\rm i}\omega))x \tilde{\cal V}_2(x)
-2{\rm i}\varepsilon \omega x(1-x)\partial_x \tilde{\cal V}_2(x) -
 p q \tilde{\cal V}_2(x) \;\;\;\;\;\;\;\;\;\;\;\;
 \;\;\;\;\;\;\;\;\;\;\; (28a)
\end{eqnarray*}
where we have substituted for $C$, $D$ and $\xi$ from equations
(20a) and (23) and added $-p q \tilde{\cal V}_2(x)$ on both sides
to make the left hand side of the equation to look like the
hypergeometric equation with the identification $$p + q = 2(\alpha +
\beta)-1 \eqno(28b)$$ 
The right hand side of the above equation
includes terms of different order in $\omega$ and therefore it is
suitable for finding the solution in the expansion of this
parameter and it is obvious that the zero-th order solutions of the
equation are the hypergeometric series provided we choose $$pq
=2\alpha \beta - j(j+1)\eqno(28c)$$
To write the zero-th order solution more explicitly we note that for $2\alpha$ a non-integer the general solution in terms of hypergeometric series could be written as follows;
$$\bar{\cal V}_2(x) = A F(p,q;2\alpha;x) + B x^{1-2\alpha} F(p-2\alpha+1, q-2\alpha+1;2-2\alpha;x)$$
hence to zeroth order in $\omega$ we have;
\begin{eqnarray*}
\lefteqn{{\cal R}_2(r) \approx ({r-il\over r+il})^{1\over2}({r-r_+ \over \varepsilon})^\alpha ({r-r_- \over \varepsilon})^\beta  {} }\\ & & {}\left(A F(p,q;2\alpha; {r_+ -r\over \varepsilon} ) + B ({r_+ -r\over \varepsilon}) ^{1-2\alpha} F(p-2\alpha+1, q-2\alpha+1;2-2\alpha; {r_+ -r\over \varepsilon} )\right) 
\;\;\;(28d)
\end{eqnarray*}
where $A$ and $B$ are constants and $p$ and $q$ are solutions of equations (28b) and (28c).

\subsection{Solution to the equation (21a)}
Amazingly enough one can see that the same procedure used to
transform equation (21b) to (28) can also be applied to equation
(21a). The reason for this is the fact that the same transformation as in
(22a) will remove the singularities of (21a) at $r = \pm {\rm i}l$,
transforming it into;
\begin{eqnarray*}
\lefteqn {\partial^2_r {\cal V}_0(r) + 2({1\over r-r_+}+
{1\over r-r_-})\partial_r {\cal V}_0(r)+
\hspace{1.5in} }\\ & & \left( {A'\over (r-r_+)^2} + {B'\over (r-r_-)^2}
+{C'\over (r-r_+)(r-r_-)} + D' ({1\over r-r_+} + {1\over r-r_-}) +
\omega^2 \right){\cal V}_0(r)=0\;\;\;\;\;\; (29)
\end{eqnarray*}
where $${\cal R}_0(r) = ({r-{\rm i}l\over r+{\rm i}l})^{1/2} {\cal
V}_0(r)\eqno(30)$$ and
$$A' = r_+ (2M\omega^2 + {\rm i}\omega) + b^2 $$
$$B' = r_- (2M\omega^2 + {\rm i}\omega) + b^2 $$
$$C' = 2M (2M\omega^2 - {\rm i}\omega) + 2(b^2+b) - {\xi^\prime}$$
$$D' = 2M\omega^2 - {\rm i}\omega \eqno(31)$$
It is notable that these equations are the same as in (23) with
${\rm i}\omega$ transformed into $-{\rm i}\omega$ and $\xi$ to ${\xi
^\prime}=\xi - 2$. Now applying the same change of the variable as
in (24) and substituting for
$${\cal V}_0 (r) =(-x)^{\alpha^\prime+1}
(1-x)^{\beta^\prime+1} \exp (i \varepsilon \omega
x)\tilde{\cal V}_0(x)\eqno(32a)$$
where
$${\alpha^\prime}_{\pm} = {1\over 2}(-3 \pm \sqrt{1-4A'})\;\;\;\;\;\;\; , \;\;\;\;\;\;\;
{\beta^\prime}_{\pm} = {1\over 2}(-3 \pm \sqrt{1-4B'})\eqno(32b)$$
equation (29) will transform into
\begin{eqnarray*}
\lefteqn {x(1-x)\partial^2_x \tilde{\cal V}_0(x)+[2{\alpha^\prime}
- 2({\alpha^\prime} + {\beta^\prime})x]\partial_x \tilde{\cal V}_0(x) -
p^{\prime} q^{\prime} \tilde{\cal V}_0(x)= \hspace{1.2in}}
\\ & & [2{\rm i}\varepsilon \omega ({\alpha^\prime}+{\beta^\prime}) +
2\varepsilon(2M\omega^2 - {\rm i}\omega)]x \tilde{\cal V}_0(x) + 2{\rm
i}\varepsilon \omega x(1-x)\partial_x \tilde{\cal V}_0(x)- p' q'\tilde{\cal
V}_0(x)+ \hspace{1.2in} \\ & &\left[(2M+\varepsilon)(2M\omega^2 - {\rm
i}\omega)+2(3l^2\omega^2+2l\omega)-j'(j'+1)+ 2({\alpha^\prime}+1)
({\beta^\prime}+1)+2{\rm i}\varepsilon \omega {\alpha^\prime} \right]
\tilde{\cal V}_0(x)\;(33)
\end{eqnarray*}
where
$$j'(j'+1) = {\xi^\prime}+ 4b^2 + 2b \eqno(34) $$
As one can see the left hand side of the above equation is also the
hypergeometric equation with the identification $p' + q' = 2({\alpha^\prime}
+ {\beta^\prime})-1$. As in equation (28a) the right hand side of this
equation includes terms of different order in $\omega$ and it is
therefore suitable for finding the solution in the expansion of
this parameter. Again the zero-th order solution is the
hypergeometric function provided we choose
$$p' q' = j'(j'+1)-2({\alpha^\prime}+1)({\beta^\prime}+1).$$
\section{Equation satisfied by $\Phi_1$ and its solution}
To complete our discussion we need to solve the equation satisfied
by scalar $\phi_1$ which, using equations (11) and (12), can be
written in the following form;
$$\left[{\cal L}_1^\dag {\cal L}_0 + \Delta ({\cal D}_1 - {r\over r^2+l^2})
({\cal D}_0^\dag +{r-2{\rm i}l\over r^2 +l^2})\right]\Phi_1(r,\theta)=0
\eqno(35)$$
 Now upon choosing the form $$\Phi_1(r,\theta) = {\cal R}_1(r)
\Theta_1$$ will transform into the following angular and radial
equations;
$${\cal L}_1^\dag {\cal L}_0 \Theta_1(\theta) = (\partial_\theta - Q +
 {\rm cot}\theta )(\partial_\theta +
Q)\Theta_1 (\theta)=-\eta \Theta_1 (\theta)\eqno(36)$$

$$\Delta ({\cal D}_1 - {r\over r^2+l^2})
({\cal D}_0^\dag + {r-2{\rm i}l\over r^2+l^2} ){\cal R}_1(r) = \eta {\cal
R}_1\eqno(37)$$ Now our task is to solve the above two equations.
\subsection{solution to the angular equation}
We start with $(36)$ which can be solved in the same way as we
solved equations (17a) and (18a). Substituting for $Q$ from its
definition and changing the variable to $x=\cos \theta$, equation
(36) will transform into;
$$\left[ (1-x^2){{\rm d}^2 \over {\rm dx}^2} - 2x {{\rm d} \over {\rm
d x}}- {(2mx + 4b^2x^2 + m^2 + 4bmx + 2bx^2 + 2b) \over
1-x^2}+\eta \right]\Theta_1(x) =0 \eqno(38)$$ where $b=l\omega$.
One can see that the only difference between this equation and
equation (19) is the disappearance of $1$ in the numerator of
$1-x^2$. In a similar way to equation (19), the dominant behavior
of the eigenfunctions of the above equation at singular points
$x=\pm 1$ are given by $(1\mp x)^{\sqrt{{\cal B}_{\pm}^2-1 }\over
2}$ where ${\cal B}_{\pm}=|1+2b \pm m|$. Introducing
$\alpha^{\prime} = ({\cal B}_{+} +1)^{1/2}({\cal B}_{+}-1)^{1/2}$
and $\beta^{\prime} = ({\cal B}_{-} +1)^{1/2}({\cal
B}_{-}-1)^{1/2}$, and substituting for $\Theta_1$ the following
expression;
$$({1-x \over 2})^{\alpha^{\prime} /2}
 ({1+x \over 2})^{\beta^{\prime} /2} U^{\prime}(x)$$
in (38) it will be transformed into;
$$(1-x^2){{\rm d}^2 U^{\prime}(x) \over {\rm dx}^2} + [\beta^{\prime} -
\alpha^{\prime} -(2+\alpha^{\prime} + \beta^{\prime})x ]{{\rm d}U^{\prime}(x)
\over {\rm dx}}+[\eta - ({\alpha^{\prime} +\beta^{\prime} \over 2})
({\alpha^{\prime}+\beta^{\prime}\over
2}+1)+4b^2+2b]U^{\prime}(x)=0\eqno(39)$$ This is again the
differential equation satisfied by the Jacobi polynomials if we
identify;
$$\eta + 4b^2 + 2b = [n+({\alpha^{\prime}+\beta^{\prime} \over 2})]
[n+({\alpha^{\prime}+\beta^{\prime} \over 2}+1)]=j(j+1)\eqno(40)$$
where $j=n+({\alpha^{\prime}+\beta^{\prime} \over 2})$.
\subsection{solution to the radial equation}
Substituting for the operators from their definitions given in (9),
the radial equation (37) can be written in the following form;
\begin{eqnarray*}
\lefteqn{{\partial^2_r {\cal R}_1(r) + 2({{r-M}\over \Delta}- {{\rm
i} l\over r^2+l^2})\partial_r {\cal R}_1(r)} + \left( {\omega^2 (r^2 +
l^2)^2\over \Delta^2}\right){\cal R}_1(r) \hspace{1.5in} }\\
\\ & &  - \left({2 \omega l
+\eta \over \Delta} -{2(r-M)(r-2{\rm i}l)\over \Delta (r^2+l^2)} -
{6{\rm i} l r +l^2-2r^2\over (r^2+l^2)^2} \right){\cal R}_1(r) = 0
\hspace{1.8in}(41)
\end{eqnarray*}
One can remove the singularity at $r=il$ by the following
transformation;
$${\cal R}_1(r) = \frac{(r-il)^{1/2}}{(r+il)^{3/2}}{\cal V}_1(r)\eqno(42)$$
so that equation (41) transforms into;
$${\partial^2_r {\cal V}_1 + 2({{r-M}\over \Delta}- {1\over
r+il})\partial_r {\cal V}_1} +\left( {\omega^2 (r^2 + l^2)^2
 \over \Delta^2}- {2 \omega l
+\eta \over \Delta}\right){\cal V}_1 =0\eqno(43)$$ 
Then by the following change of variable,
$${r - r_{+}\over 2M} = -z$$ 
the above equation transforms into
$$ \partial^2_z {\cal V}_1 + \left( {1\over z} + {1\over z-1} - {2\over z-a}\right)\partial_z {\cal V}_1 + {\alpha\beta z - q \over z(z-1)(z-a)}{\cal V}_1 = {\omega^2 \over {2M}^3}{[(r_+ - 2Mz)^2 + l^2]^2 \over {z^2 (z-1)^2}}{\cal V}_1\eqno(44a)$$
where
$$\alpha + \beta = -1 \;\; ; \;\; \alpha\beta={1\over 2M}(2\omega l + \eta)\eqno(44b)$$ 
and 
$$q= \alpha \beta a\;\;\; ; \;\;\; a={r_+ + {\rm i}l\over 2M}\eqno(44c)$$ 
Comparing equations (44a-c) with Heun's differential equation [17] defined by 
$${d^2y \over dz^2} + \left( {\gamma \over z} + {\delta \over z-1} + 
{\epsilon \over z-a}\right){dy \over dz} + {{\alpha\beta z - q}\over {z(z-1)(z-a)}}y = 0\eqno(45)$$
 where 
 $$a \neq 0, 1 \;\; {\rm and} \;\; \gamma + \delta + \epsilon = \alpha + \beta + 1$$ we find out   
that equation (44a), along with the fact that $|a|>1$, forms the Heun equation with the right hand side including a term of second order in $\omega$, allowing solutions in the expansion of this parameter. Again the zeroth order solutions are solutions to the Heun equation but before looking for that the following two points worth to be mentioned;\\
First we note that despite the fact that $q=a\alpha\beta$, the left hand side of 
equation (44a) does not degenerate into hypergeometric equation because $\epsilon \neq 0$ and
secondly, due to the fact that all the parameters $\gamma ,\; \delta \; {\rm and} \; \epsilon$
are integers we expect that the general solution to the Heun equation involve logarithmic terms.\\ 
To write the explicit zero-th order solution to the equation (44a) we only consider the basic power series solution \footnote{ For detailed dicussion on possible different solutions to Heun's equation see [18-19] } to the Heun equation denoted by [18];
$$Hl(a,q;\alpha,\beta,\gamma,\delta;z)= \sum_{j=0}^\infty c_j z^j\;\;\;\; , \;\;\;\; (c_0=1)\eqno(46)$$
where $Hl$ stands for 'Heun local' solution and apart from the normalized coefficint 
$c_0$ the other coefficints are given by the following three-term recursion relation;
$$-qc_0 + a\gamma c_1 = 0 \eqno(47)$$
$$P_j c_{j-1} - (Q_j + q) c_j + R_j c_{j+1} =0 \;\; , \;\;(j\geq 1) \eqno(48)$$
in which ;
$$P_j = (j-1+\alpha)(j-2-\alpha) \eqno(49a)$$
$$Q_j = j[j(1+a)+a-2] \eqno(49b)$$
$$R_j = a(j+1)^2 \eqno(49c)$$
where we used relations (44b) and the fact that in our case, left hand side of equation (44a), $$\gamma = \delta =1\;\;\;\; {\rm and}\;\;\;\; \epsilon=-2$$
Therefore the zero-th order solution to the equation (44a) is given by;
$${\cal R}_1(r) = \frac{(r-il)^{1/2}}{(r+il)^{3/2}}Hl(a,q;\alpha,\beta,1,1;{r_{+} - r\over 2M})\eqno(50)$$

It is also interesting to note that Heun's equation and its biconfluent extension also appear in the discussions on the electromagnetic perturbations of Kerr-de Sitter and Kasner spacetimes respectively
and indeed furnish exact solutions to the Maxwell equations in these spacetimes [20,21].

\section{Results and Discussion}
In this paper we have considered the Maxwell equations in NUT
space and solved them analytically using the Newman-Penrose null
tetrad formalism, which is best adapted for treating type-D
spacetimes. It is shown that after separation of the equations,
solutions to the angular equations could be given in terms of the
Jacobi polynomials. The radial equations on the other hand are
transformed into hypergeometric and Heun's equations which on the 
right hand side have terms of different order in $\omega$. This 
 fact enables one
to find solutions in the expansion of this parameter with the
zero-th order solutions being the hypergeometric function and possible 
solutions to the Heun equation respectively.
To compare the results with the Schwarzscild case we note that the appearance of the perturbation frequency $\omega$ in the angular equations (19) and (38) 
, only through the parameter $b=l\omega$, means that unlike Schwarzschild, in the stability discussions of NUT space, we can not only consider radial equations and angular equations should also be considered. In this respect the behaviour of NUT space under electromagnetic pertubations is very similar to that of Kerr in which angular equations depend on the frequency through the combination $a \omega$ [11] . So in both cases going over to the Schwarzschild space will remove the perturbation parameter from the angular equation and the stability discussion will only need a study of radial equations.
Another point deserving attention is to check whether, as in 
the case of the Kerr geometry, it is possible to write
the solutions to the radial equations in the form of an expansion
in terms of the coulomb wave functions [17,22].
\section*{acknowledgements}
I would like to thank D. Lynden-Bell for his valuable comments on
the first draft of the paper and J. Bicak for useful discussions.
The author thanks University of Tehran for supporting this project 
under the grants provided by the Research Council.


\begin{references}
\bibitem{1} E. T. Newman, L.Tamburino and T. Unti, J. Math. Phys.,
 {\bf 4}, 915, 1963.
\bibitem{2} C. W. Misner, J. Math. Phys. {\bf 4}, 924, 1963.
\bibitem{3} Gibbons, G. W., Manton, N. S. (1986) Nuclear Physics {\bf B274}, 183.
\bibitem{4} T. C. Kraan , P. Baal (1998),  INLO-PUB-4/98, hep-th/9802049.
\bibitem{5} K. Behrndt, From black holes to D-Branes, Humboldt University Lectures, Berlin, 1996-1997.
\bibitem{6} M. J. Duff, R. R. Khuri and J. X. Lu, Physics reports {\bf 259} (1995), 213-326 hep-th/9412184.
\bibitem{7} D. Lynden-Bell and M. Nouri-Zonoz, Rev. Mod. Phys., {\bf
70}, 427, 1998.
\bibitem{8} D. Bini, C. Cherubini, R.T. Jantzen and B. Mashhoon, 
Phys.Rev. D67 (2003) 084013, gr-qc/0301080.
\bibitem{9} B. Mukhopadhyay and N. Dadhich, gr-qc/0301104.
\bibitem{10} M. Nouri-Zonoz and A. R. Tavanfar, JHEP02(2003)059, hep-th/0212325.
\bibitem{11} S. Chandrasekhar, The mathematical theory of black
holes, OUP, 1983.
\bibitem{12} E. T. Newman and R. Penrose, J. Math. Phys.,
 {\bf 4}, 566-79, (1962).
\bibitem{13} R. Geroch, A. Held and R. Penrose, J. Math. Phys.,
 {\bf 14}, 874-81, (1973).
\bibitem{14} S. A. Teukolsky, Phys. Rev. Lett., 29, 1114-18, 1972; Astrophys. J., 185, 635-49, 1973.
\bibitem{15} I. S. Gradshteyn  and I. M. Ryzhik, Tables of integrals, series and products, 5-th edition, Academic press, 1994.
\bibitem{16}E. D. Fackerell and R. G. Crossman, J. Math. Phys., {\bf 18}, No. 9, 1977.
\bibitem{17} S. Mano, H. Suzuki and E. Takasugi, Prog. Theor. Phys. 95 
(1996) 1079-1096, gr-qc/9603020.
\bibitem{18} Heun's differential equation, Ronveaux A., Ed., OUP, 1995. 
\bibitem{19} H. Suzuki, E. Takasugi and H. Umetsu, Prog. Theor. Phys. 100 
(1998) 491-505, gr-qc/9805064.
\bibitem{20} H. Suzuki, E. Takasugu and H. Umetsu, Prog. Theor. Phys. 102 (1999) 253-272, gr-qc/9905040.
\bibitem{21} R. Pons and G. Marcilhacy, Class. Quantum Grav. {\bf 4} (1987)
 171-179.
 \bibitem{22} E. W. Leaver, J. Math. Phys., {\bf 27},  No. 5, 1986.\\
\end{references}
\end{document}